\documentstyle[12pt,aps]{revtex}
\begin{document}

\title{
\begin{flushright}
\vspace{-1cm}
{\normalsize MC/TH 96/09}
\vspace{1cm}
\end{flushright}
\bigskip}
\author{Michael C. Birse}
\address{Theoretical Physics Group, Department of Physics and Astronomy\\
University of Manchester, Manchester, M13 9PL, UK\\}
\maketitle
\bigskip
\bigskip
\begin{abstract}
The contribution of nucleons to the quark condensate in nuclear matter includes
a piece of first order in $m_\pi$, arising from the contribution of 
low-momentum virtual pions to the $\pi N$ sigma commutator. Chiral symmetry
requires that no term of this order appears in the $NN$ interaction. The mass
of a nucleon in matter thus cannot depend in any simple way on the quark
condensate alone. More generally, pieces of the quark condensate that arise
from low-momentum pions should not be associated with partial restoration of
chiral symmetry.

\end{abstract}
\bigskip

There has been much recent interest in the question of whether chiral symmetry
is partially restored in nuclear matter.\footnote{For reviews of this topic
and further references, see\cite{abrev,birrev,brrev}.} The chiral isospin
symmetry SU(2)$\times$SU(2) is an approximate symmetry of the QCD Lagrangian,
broken only by the small current masses of the up and down quarks. This
symmetry is realised in the hidden (spontaneously broken) mode since the QCD
vacuum is not invariant under it. The scalar density of quarks, often referred
to as the ``quark condensate," provides an order parameter describing the
hidden symmetry. Any reduction of this condensate in matter has generally been
interpreted as a signal of partial restoration of the symmetry.

The quark condensate in vacuum can be related to pion properties through the
Gell-Mann--Oakes--Renner (GOR) relation,
$$m_\pi^2 f_\pi^2\simeq -2\bar m\langle \overline qq\rangle_0, \eqno(1)$$
where $\bar m$ is the average of the current masses of the up and down quarks
and $\langle \overline qq\rangle_0$ is the quark condensate (per quark
flavour) in vacuum. The change in this condensate in matter, to first order in
the density, can be estimated in a model-independent way from\cite{dl,cfg,lkw}
$${\langle \overline qq\rangle_\rho\over\langle \overline qq\rangle_0}
=1-{\sigma_{\pi{\scriptscriptstyle N}}\over f_\pi^2 m_\pi^2}\rho,
\eqno(2)$$
where $\sigma_{\pi{\scriptscriptstyle N}}$ 
is the pion-nucleon sigma commutator,
$$\sigma_{\pi{\scriptscriptstyle N}}=2\bar m\langle N|\int d^3\hbox{r}\,
\overline qq|N\rangle. \eqno(3)$$
The current best determination of $\sigma_{\pi{\scriptscriptstyle N}}$ from
$\pi N$ scattering is $45\pm 7$ MeV\cite{gls}. Using this in (2) suggests a
$\sim 30\%$ reduction in the quark condensate at normal nuclear matter
densities.

There has been some debate about corrections to this result\cite{cfg,me,bm}. 
Estimates using both simple models\cite{cfg,ce,bm} and relativistic BHF 
calculations with realistic $NN$ forces\cite{lk,bw} suggest that corrections
are small up to normal nuclear densities, although they rapidly become large
(and highly model-dependent) beyond that. There have also been questions about
whether the effect is sufficiently short-ranged that the hard-core correlations
between nucleons mean that one nucleon does not feel the change in the
condensate produced by its neighbours\cite{te}. However, the strong coupling
of a scalar meson, which can be interpreted as an excitation of the
condensate, to two-pion states suggests that the effect should be long-ranged
and so not cut out by the correlations\cite{bir}.

In contrast, there has been little discussion of whether changes in the quark
condensate necessarily mean partial symmetry restoration, or indeed if those 
changes can have directly observable effects on particle properties in matter.
It is these questions that I address here.

An indication that $\langle \overline qq\rangle_\rho$ may not be the most
relevant quantity is provided by the chiral expansion of the energy of a
nucleon in nuclear matter. To first order in the density the mass of a nucleon 
in matter (at mean-field level) can be written in the form
$$M_{\scriptscriptstyle N}^*=M_{\scriptscriptstyle N}+V_s(0)\rho, 
\eqno(4)$$
where $V_s(q^2)$ is the part of the $NN$ potential (or more specifically, the 
two-nucleon irreducible $NN$ scattering amplitude) with a scalar-exchange
character.\footnote{This definition corresponds to a mass that could appear
in a Dirac equation for the nucleon. The irreducible $NN$ amplitude in (4)
avoids the problem of the strong energy dependence of the full scattering
amplitude produced by the deuteron pole just below threshold, discussed
in\cite{fh}. The irreducible amplitude is also the one to which Weinberg's
chiral counting rules apply\cite{wein}.} Suppose, for the sake of argument,
that the mass in matter contained a piece that depended in some way on the
quark condensate so that, to first order in the density, there was a term in
$M_{\scriptscriptstyle N}^*$ proportional to $\sigma_{\pi{\scriptscriptstyle
N}}\rho/(f_\pi^2 m_\pi^2)$. This would imply the existence of a scalar piece
of the $NN$ interaction that was proportional to $\sigma_{\pi N}/m_\pi^2$ at
zero momentum transfer\cite{bir}.

The chiral expansion of $\sigma_{\pi N}$ has the form\cite{gl}
$$\sigma_{\pi{\scriptscriptstyle N}}=Am_\pi^2-{9\over 16\pi}\left({g_{\pi 
{\scriptscriptstyle NN}}\over 2 M_{\scriptscriptstyle N}}\right)^2m_\pi^3
+\cdots, \eqno(5)$$
where the constant $A$ involves a counterterm corresponding to short-distance
physics, whose value must be determined empirically from $\pi N$ scattering.
In contrast, the ${\cal O}(m_\pi^3)$ term is nonanalytic in the current quark
mass and so arises purely from the long-distance physics of the pion cloud
surrounding a nucleon. Its coefficient is given entirely in terms of the $\pi
N$ coupling and nucleon mass. If there were a piece of the $NN$ interaction
proportional to $\sigma_{\pi N}/m_\pi^2$ (with a coefficient independent
of $m_\pi$), it would contain a corresponding term of order $m_\pi$. However
Weinberg's dimensional counting rules\cite{wein}\footnote{For a review
see\cite{bkm}.} show that no such term can be present in the chiral expansion
of the $NN$ potential. Chiral symmetry thus rules out any interaction
proportional to $\sigma_{\pi{\scriptscriptstyle N}}/m_\pi^2$ on its own.
Either there are additional terms that cancel out the piece of order $m_\pi$,
or the coefficient of $\sigma_{\pi{\scriptscriptstyle N}}/m_\pi^2$ is at least
of order $m_\pi^2$ (in which case the term could make only a small
contribution to the full scalar-exchange interaction). Hence there can be no
direct dependence of the nucleon mass in matter on the quark condensate (2).

As an illustration of this point, consider the calculations in Ref.\cite{bir}
using the linear sigma model. There I defined a ``symmetry restoring
amplitude" as a sum of $NN$ diagrams in which the $\sigma$ meson coupled
directly to one of the nucleons. This has precisely the form just discussed,
proportional to $\sigma_{\pi {\scriptscriptstyle N}}/m_\pi^2$. However as
pointed out in\cite{bir}, the full $NN$ scattering amplitude arising from
exchange of two virtual pions involves a different combination of
diagrams. In these there is a strong cancellation between diagrams in which
the pions couple to a nucleon via a $\sigma$ and others in which the pions
couple directly, just like the ``pair suppression" in $\pi N$ scattering.
These cancellations ensure that the ${\cal O}(m_\pi)$ term arising from the
sigma commutator does not appear in any chirally symmetric calculation of
the full $NN$ amplitude\cite{nnch,bir}.

In general, the contributions to the quark condensate from low-momentum pions
need to be distinguished from those of other hadrons because of the special
role pions play as approximate Goldstone bosons. The integrated scalar density
of quarks in a pion can be estimated by applying the Feynman-Hellmann theorem
to the pion mass and using the fact that $m_\pi^2\propto \bar m$, as in (1).
This leads to
$$2\bar m\langle \pi|\int d^3\hbox{r}\,\overline qq|\pi\rangle
=m_\pi^2{d m_\pi\over d m_\pi^2}={\textstyle {1\over 2}}m_\pi
\simeq 70\ \hbox{MeV}. \eqno(6)$$
This is even larger than the corresponding matrix element for the nucleon (3).
Hence the scalar density of quarks in a pion is large, reflecting the
collective nature of a Goldstone boson. Note that (6) applies to a pion
state that has been unconventionally normalised to unity. For a more familiar
covariantly normalised pion, the corresponding matrix element is $m_\pi^2$. If
one naively assumed that the scattering of a nucleon from a pion contained a
term proportional to the scalar quark density of the pion, then one would
expect this to give a contribution to $\pi N$ scattering of order $m_\pi^0$.
However it has long been known that this is not the case. Chiral symmetry
requires that the isospin-averaged amplitude for pion-nucleon scattering is of
order $m_\pi^2$ at threshold\cite{wpin}. The contribution of these pions to
observables such as scattering amplitudes or the masses of nucleons and other
hadrons (except pions and kaons) in matter is thus suppressed by a factor of
$m_\pi^2$ compared with their quark density.

In the particular case of the $NN$ interaction, the leading nonanalytic
dependence on the current quark mass, the ${\cal O}(m_\pi^3)$ term, in
$\sigma_{\pi{\scriptscriptstyle N}}$ arises from the longest-ranged part of
the pion cloud, corresponding to virtual pions of very low momentum. These
pions are responsible for the ${\cal O}(m_\pi)$ piece of the scalar density of
the nucleon but, as in the real $\pi N$ scattering just discussed, their
contribution to the scattering of another nucleon is suppressed by a factor of
$m_\pi^2$. This can also be seen by applying the counting rules\cite{wein} to
a two-pion-exchange diagram with one $NN\pi\pi$ vertex of dimension $d=2$. From
this one finds a contribution to the $NN$ interaction of order $m_\pi^3$.
Hence, as we have already seen, the part of (2) which is of order $m_\pi$
cannot contribute a piece of the same order to the energy of a nucleon in
matter.

Note that short-ranged correlations between nucleons, which can also destroy
any simple dependence of the nucleon mass on the quark condensate in 
matter\cite{te,dce}, are not relevant to the question here. Since the parts of
the $NN$ interaction under discussion arise from long-ranged pion exchanges,
they are not affected by such short-range correlations.

Analogous effects can be found in the properties of nucleons in vacuum. The
quark condensate in vacuum includes a piece of order $m_\pi^2\ln m_\pi$,
also generated by virtual pions\cite{lp}. However logarithmic terms in the
chiral expansion of the nucleon mass start at order $m_\pi^4\ln
m_\pi$\cite{gss}. In any model where the nucleon mass contains a term
proportional to the quark condensate, there are again strong cancellations
of that with other pion cloud contributions, as illustrated by the one-loop
calculation of $\sigma_{\pi{\scriptscriptstyle N}}$ in the linear sigma
model\cite{bir}.

The QCD sum rule approach\cite{svz} to hadron properties involves a
short-distance operator product expansion (OPE). This brings in matrix
elements of local operators like $\overline qq$. These cannot discriminate
between contributions from low-momentum virtual pions and those from other
components of the QCD vacuum. The OPE side of a sum rule for the nucleon
mass\cite{iof} thus contains a piece of order $m_\pi^2\ln m_\pi$\cite{gc}. A
careful treatment of the continuum, including low-momentum $\pi N$ states, is
needed on the phenomenological side of the sum rule to obtain an expression
for the mass without the offending logarithm\cite{lccg}.

Matter at finite temperature also demonstrates similar effects. At low 
temperatures and zero baryon density this is a pion gas. The scalar quark 
density of this produces a temperature dependence of the quark condensate
that starts at order $T^2$\cite{gel}. In contrast the change in the mass of 
a nucleon in this gas starts at order $T^4$\cite{ls}. In finite-temperature
QCD sum rules for the nucleon the ${\cal O}(T^2)$ dependence of the condensate
is cancelled by including the $\pi N$ scattering on the phenomenological side
of the sum rule\cite{koi}.

A proper treatment of the $\pi N$ continuum in matter would presumably remove
the corresponding ${\cal O}(m_\pi)$ term that arises on the OPE side of sum
rules for a nucleon in matter\cite{nmsr}. Until this is done, the
corresponding piece of $\langle \overline qq\rangle_\rho$ should be regarded
as an additional source of uncertainty. Its size can be estimated from the
${\cal O}(m_\pi^3)$ term in $\sigma_{\pi{\scriptscriptstyle N}}$, which is 
$-25$ MeV and so is about half of the coefficient of the density in (2).
This presence of this piece in the sum rules for a nucleon in matter should be
regarded as a significant extra uncertainty in their results.

If these contributions from low-momentum pions to changes in the quark
condensate do not affect the masses of other hadrons, one should also ask
whether they correspond to restoration of chiral symmetry. In the framework of
a sigma model, one can imagine large-amplitude, low-momentum fluctuations of
the pion fields around the chiral circle (or indeed a pion
condensate\cite{bkr}) that could significantly reduce the mean value of the
$\sigma$ field without moving the fields off the chiral circle
$\sigma(x)^2+\hbox{\boldmath$\pi$}(x)^2=f_\pi^2$. Since the $\sigma$ field
corresponds to the scalar quark density of the underlying QCD description, one
might be tempted to conclude from the its reduction that chiral symmetry has
been partially restored. However the fact that fields remain on the chiral
circle indicates that this is not the case. Similarly a reduction in $\langle
\overline qq\rangle$ from its vacuum value is not enough for one to conclude
that the symmetry has been partially restored, if part of this change arises
from low-momentum pions.

As a corollary to this, one can imagine situations where the quark condensate
vanishes but chiral symmetry is not restored. For example, this could happen
if the linear density dependence of (2) were to continue to hold up to around
three times the density of nuclear matter, although that seems unlikely from
present estimates of higher-order density dependence\cite{lk,bw}. The pion
would then be anomalously light, in the sense described in\cite{cb}. The
leading term in the GOR relation would vanish and the pion mass would be
proportional to the current quark mass $\bar m$ rather than its square
root.\footnote{There is a school of chiral perturbation theory that suggests
that this is already the situation in the normal vacuum. See, for
example\cite{scad}.} This could indicate proximity to a pion-condensed phase
of the type described in\cite{bkr}. In any case other condensates can play an
increasingly important role as $\langle \overline qq\rangle$ tends to zero.

In summary: the quark condensate in matter contains pieces arising from
low-momentum virtual pions. These do not contribute to nucleon properties in
matter and cannot be associated with a partial restoration of chiral symmetry.

\section*{Acknowledgements}

I am grateful to B. Krippa for the question that prompted this work, and to him
and J. McGovern for their comments on it. This work was supported by the EPSRC 
and PPARC.

\end{document}